\documentclass[apl,twocolumn,showpacs,preprintnumbers,amsmath,amssymb]{revtex4-1}

\usepackage{graphicx}
\usepackage{dcolumn}
\usepackage{bm}
\usepackage{textcomp}
\usepackage{wasysym}

\begin{document}

\preprint{APS/PRB/2DMembranes}


\title{Zero-field thermopower of a thin heterostructure membrane with a 2D electron gas}

\author{M.\ Schmidt, G.\ Schneider, Ch.\ Heyn, A.\ Stemmann, and W.\ Hansen}
\affiliation{Institut f\"ur Angewandte Physik und Zentrum f\"ur
Mikrostrukturforschung,\\Jungiusstra\ss e 11, D-20355 Hamburg,
Germany}

\date{\today}

\begin{abstract}
We study the low-temperature thermopower of micron sized, free-standing membranes containing a two-dimensional electron system. Suspended membranes of 320 nm thickness including a high electron mobility structure in Hall bar geometry of 34 $\mu$m length are prepared from GaAs/AlGaAs heterostructures grown by molecular beam epitaxy. Joule heating on the central region of the membrane generates a thermal gradient with respect to the suspension points where the membrane is attached to cold reservoirs. Temperature measurements on the membrane reveal strong thermal gradients due to the low thermal conductivity. We measure the zero-field thermopower and find that the phonon-drag contribution is suppressed at low temperatures up to 7 K. 
\end{abstract}


\maketitle

\section{Introduction} 
\begin{figure}[t]
\centering
\includegraphics[width=8.5cm]{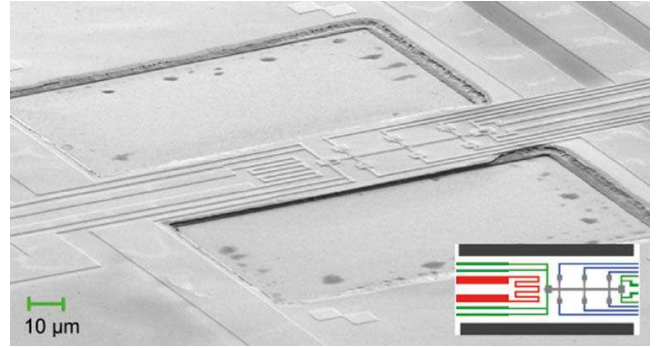}
\caption {(color online) Scanning electron micrograph of a suspended, 320 nm thick, 40 $\mu$m wide, and 100 $\mu$m long GaAs membrane with a Hall bar device composed of a GaAs/AlGaAs heterostructure containing a 2DEG. The inset shows a schematic drawing of the membrane with color coded features. A micro heater (red) and thermometers (green) enable the generation and the measurement of the temperature drop directed along the Hall-bar (gray), respectively.}
\label{fig:SEM-image}
\end{figure}

Nanowires and thin membranes have attracted much attention in the field of thermoelectrics because their thermal conductivities are several orders of magnitude reduced compared to bulk material. \cite{Carlson1965.jap, Holland.1963.prev., Fon2002.prb} There are several publications on the thermal properties of micro-scaled structures as for example GaAs microbars \cite{Fon2002.prb, Tighe1997.apl} and Si nanowires.\cite{Hochbaum2008.nature, Boukai2008.nature} In recent years, many reports have been published on the thermopower of such nanowires and it became common practice to prepare the nanowires in a free-standing  fashion\cite{Hochbaum2008.nature} to decouple the wires from the substrate avoiding heat losses to or influences of the latter. Thermopower studies on two-dimensional electron gases (2DEGs) in heterostructures that have been performed on bulk substrates, still show strong interactions of the electronic system with the substrate. \cite{basic-prop-of-semiconductors,Ying1994.prl,Fletcher1995.prb}  Here we report about thermopower studies of suspended 2DEGs.\\
Thermoelectric studies of two-dimensional electron systems have started shortly after the discovery of the quantum Hall effect \cite{Obloh.1984.surfscience} and have attracted considerable interest  since then. \cite{basic-prop-of-semiconductors, Fromhold1993.prb, Fletcher1994.prb, Cyca1992.JPCondMatter, Sankeshwar.pss, lee2009.apl, Tieke1995.prb}  Generally, a thermal gradient established by Joule-heating along a macro scaled sample is used to investigate the thermopower associated with the temperature drop between the voltage probes. Then, two effects are expected to contribute to the thermopower, the thermo-diffusion and the phonon drag. It is of interest how their influence might be modified in suspended nanostructures. Phonon-drag thermopower is found to dominate in a temperature range that starts at very low temperature, i. e., below $T$ = 1 K in bulk GaAs.\cite{basic-prop-of-semiconductors, Ying1994.prl} Different efforts were made to experimentally separate the diffusion thermopower and the phonon-drag contribution. Ying et al. \cite{Ying1994.prl} used a GaAs substrate that was thinned to 100 $\mu$m. They were able to measure the pure thermo-diffusion of a 2D hole gas at temperatures below 100 mK. Fletcher et al. \cite{Fletcher1995.prb} used a heavily doped substrate and suppressed the phonon drag up to 0.5 K. Finally, the use of direct heating techniques of the two-dimensional electron gas reduces the thermal gradient in the lattice and suppresses the phonon drag in the electronic system. This way it has been possible to study the diffusion thermopower up to a temperature of $T$ = 2 K, which is the highest value reported for GaAs yet.\cite{Maximov2004.prb, Chickering2009.PRL}\\
We present thermopower measurements on suspended micro-scaled 2DEGs confined in a thin GaAs/AlGaAs heterostructure membrane as shown in Fig. \ref{fig:SEM-image}.
The sample setup combines the advantageous properties of drastically reduced heat conductivity of free-standing, thin membranes and the high electric conductivity of 2DEGs. While electrical transport properties of 2DEGs embedded in thin membranes have been studied in several publications \cite{Blick2000.prb, Mendach2004.PhysicaE, Mendach2006.apl, Friedland2007.prb},  thermal transport is rarely reported and thermopower measurements are missing, so far. We demonstrate that in our membranes the thermal transport of the lattice is strongly reduced due to the small dimension. This allows us to establish strong thermal gradients along the 2DEG even at distances of only several micrometers, which is not possible on structures in contact with bulk GaAs. \cite{Carlson1965.jap, Holland.1963.prev.} With our structure we observe that the small dimensions strongly affect the thermopower suppressing the phonon drag in the 2DEG up to temperatures of 7 K. 

\section{Experiment}
The measurements were carried out on a free-standing membrane containing an electron system in a high electron mobility (HEMT) structure that is shown in Fig. \ref{fig:SEM-image}.
The thickness of the membrane is 320 nm and the length is 100 $\mu$m. The lateral dimension of the membrane is about two orders of magnitude smaller than samples used in previous reports of thermoelectric studies on 2DEGs in GaAs heterostructures.\cite{basic-prop-of-semiconductors, Obloh.1984.surfscience, Vuong1986.solstatcomm, Fletcher1994.prb}\\
The samples were fabricated using solid-source molecular beam epitaxy (MBE) on (001) GaAs substrates. 
A 500 nm thick Al$_{0.61}$Ga$_{0.39}$As sacrificial layer is followed by the HEMT structure with a 350 nm layer of intrinsic GaAs, a 30 nm Al$_{0.35}$Ga$_{0.65}$As spacer layer, a 57 nm Si:Ga$_{0.65}$Al$_{0.35}$As doping layer, and a 5 nm GaAs cap. A 2DEG  is formed at the GaAs / Al$_{0.35}$Ga$_{0.65}$As interface.\\
We used e-beam lithography to structure a 34 $\mu$m long micro Hall bar, a metal heater placed in the center of the membrane for Joule-heating of the crystal lattice, and two thermometers at the ends of the Hall bar as shown in Fig. \ref{fig:SEM-image}. Joule-heating of the wound heater generates a temperature gradient directed along the membrane. A 120 nm etching step defines the Hall bar containing the 2DEG and leaves the rest of the membrane semi-insulating. All leads as well as the thermometers and the heater were fabricated in one evaporation step, so that all metallizations can be assumed to have the same specific resistivity. 
From the specific resistivity we determine the resistance of the wound heating wire in order to calculate the heater power from the heater current. In our studies AuGe with a small content of $\approx 5\%$ Ni was used to benefit from the Kondo effect enhancing the sensitivity of the thermometers at low temperatures. We also tested pure Au as a metal with significantly lower specific resistivity compared to AuGeNi and could exclude any influence of the chosen metal on the measured thermopower. The film thickness of the evaporated metal was 40 nm. Finally, chemical wet etching of the mesa with H$_3$PO$_4$/H$_2$O$_2$/H$_2$O and selective wet etching of the Al$_{0.61}$Ga$_{0.39}$As sacrificial layer with a 5\% solution of HF detaches the membrane from the substrate. In the suspended membrane we determine a carrier density of the 2DEG of $n_s=1.3\cdot 10^{11}$ cm$^{-2}$. The mobility of the 2DEG in the detached membrane is about 97,000 cm$^2$/Vs. \\
The thermometers at the hot and cold end of the Hall bar are designed in four-point geometry. Both thermometers were calibrated at equilibrium condition in a slow bath-temperature sweep. The calibration curves for both thermometers are shown in the inset of Fig. \ref{fig:thermal-conductivity}. Whereas in the thermometers made of pure Au the saturation of the resistivity does not allow for temperature measurements below 10 K, the Kondo effect in the AuGeNi thermometers allows for temperature measurements down to the lowest temperature of 1.3 K available in our cryostat. \\
Both thermometers act in addition as ohmic contacts to the 2DEG by the use of segregated NiAuGe, so we can guarantee that the position at which the thermovoltage is measured can be associated with a precise temperature at the membrane. Furthermore, the segregated NiAuGe thermally anchors the electron temperature to the crystal lattice. 

\section{Thermal conductivity} \label{thermal-cond}
\begin{figure}[bt]
\centering
\includegraphics[width=0.5 \textwidth]{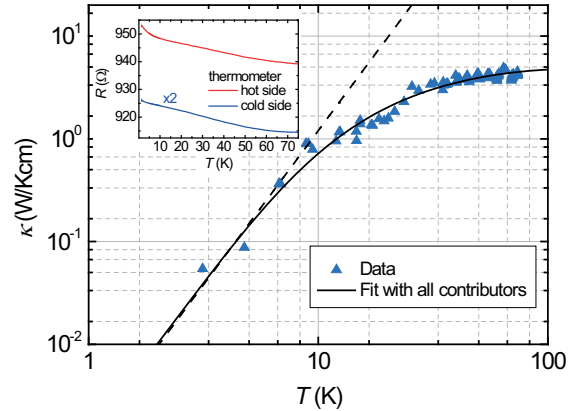}
\caption{(color online) Thermal conductivity $\kappa$  of a membrane with 42 $\mu$m x 320 nm cross sectional area. The solid line is calculated via the Callaway model. The dashed line is a function prop. to $T^3$ demonstrating the influence of the phonon specific heat at low temperatures. The inset shows the run of the thermometer resistances of the hot (red) and cold (blue, multiplied by 2 for better comparison) end side thermometer, respectively which we used for calibration.}
\label{fig:thermal-conductivity}
\end{figure}
The measurements were performed using a variable-temperature cryostat. For the measurement of the thermal conductivity, we apply Joule heating to the heater and measure the temperature drop between the thermometers. As the suspension points of the membrane act as heat sinks, a thermal gradient is established along the membrane. When a steady state is achieved, the heat flow along the membrane causes a constant temperature gradient between the thermometers and the thermal conductivity of the membrane can be determined by the temperature difference $\Delta T$ of the thermometers and the heater power $P$ as
\begin{align}
	\kappa= \frac{P}{\Delta T} \frac{\Delta x}{A},
	\label{kappa}
\end{align}
where $\Delta x$ is the distance between the thermometers and $A$ is the cross-sectional area of the membrane. Here we neglect radiation loss which is extremely small at the temperatures of our measurement. The heating power $P$ is calculated assuming a temperature independent heater resistivity of 5 k$\Omega$. This assumption is associated with a small error below 2 \% because of the small temperature dependence of the metal. \\   
We determined the thermal conductivity in a temperature range between 3 K and 75 K as shown in Fig. \ref{fig:thermal-conductivity}. The run of the thermal conductivity can be well described by a model developed by Callaway.\cite{Callaway1959.pr} We use this model with later corrections that can, e.g., be found in the publication of Fon et al.\cite{Fon2002.prb} and investigate the effects of phonon-boundary, -electron, -defect and -Umklapp scattering on the thermal conductivity of our membrane. The different phonon scattering  mechanisms enter into the model with their corresponding relaxation times , which are combined in the total relaxation time $\tau$ according to Matthiessen's rule:
\begin{eqnarray}
	\tau^{-1}&=& \tau^{-1}_{bound}+\tau^{-1}_{electron}+\tau^{-1}_{defect}+\tau^{-1}_{phonon}\\
	&=&c_{av}/\Lambda_{bound}+\alpha\nu+\beta\nu^4+\gamma\nu^2Te^{-(\Theta_D/3T)} \nonumber
	\label{eqn:tau}
\end{eqnarray}
with the phonon frequency $\nu$, the Debye temperature $\Theta_D$ and the average phonon group velocity $c_{av}\approx 3500$ m/s. Fitting parameters are $\alpha$, $\beta$, $\gamma$ and $\Lambda_{bound}$. The latter represents the limiting phonon mean free path for diffusive boundary scattering. The linear form of the phonon-electron scattering rate is adequate for scattering in a degenerate semiconductor \cite{Ziman1960} and the fourth-power dependence of the defect scattering rate is commonly used for Rayleigh scattering at point defects.\cite{Holland1964.Prev} \\
Thus, from a fit with the Callaway model, we obtain information about the rates of different phonon scattering mechanisms and their contribution to the thermal resistivity of the membrane. The parameters extracted from the fit in Fig. \ref{fig:thermal-conductivity} are $\Lambda_{bound}=67.24$~$\mu$m, $\alpha=6.68\cdot10^{-28}$, $\beta=2.43\cdot10^{-41}$~s$^3$ and $\gamma=1.73\cdot10^{-18}$~s/K.
In comparison to bulk values, where $\Lambda_{bound}$ is usually in the order of millimeters, diffusive phonon-boundary scattering in the membrane is strongly enhanced.\cite{Carlson1965.jap, Holland.1963.prev.} This can be easily understood in view of the drastically reduced cross sectional area of the thin membrane. The dashed line in Fig. \ref{fig:thermal-conductivity} is a function proportional to $T^3$ which reflects the thermal conductivity as expected due to the Debye phonon heat capacity at low temperatures. At these temperatures only boundary scattering gives the limiting factor to $\kappa$ so that the phonon mean free path can be assumed to be constant. \\
Furthermore, according to the fit results, phonon-electron scattering gives a negligible contribution to the thermal resistivity. This can be explained with the membrane setup, that contains free charge carriers only in the spatially confined plane of the 2DEG in combination with the dominance of boundary and defect scattering in the structure.\\ 
We find that the value of $\beta$ for defect scattering deduced from the fit is in the order of those measured in bulk GaAs, indicating that the rate of point defect scattering is not significantly changed in our membrane.\cite{Holland1964.Prev} Phonon-defect scattering is responsible for the reduced rise of $\kappa$ at temperatures above 10 K.\\
The value of $\gamma$ for phonon-phonon Umklapp scattering in our membrane is about an order of magnitude smaller than bulk values. We expect from the result of the Callaway model that phonon-Umklapp scattering will gain influence in the membrane not below 100 K. This is in contrast to bulk GaAs where Umklapp scattering processes already become dominant at temperatures above 10 K.\cite{Carlson1965.jap,Holland.1963.prev.}\\
We note that the run of the thermal conductivity of our membrane is comparable to thermal conductivities determined in earlier work by Fon et al. \cite{Fon2002.prb} on GaAs nanobars of 200 nm by 250 nm cross sectional area. The thermal conductivities for those membranes are two orders of magnitude reduced, which we attribute to the two orders of magnitude larger  width of our membrane with respect to the nanobars studied by Fon. \\  
In addition to the phonons, the electrons in the leads and the Hall bar contribute to the thermal conductivity $\kappa$ of the membrane. This contribution can be estimated via the Wiedemann Franz law as $\kappa_{WF,lead}=L_0\sigma_{lead} T$ and $\kappa_{WF,2DEG}=L_0\sigma_{2DEG} T$ with the Lorentz number $L_0$ and the specific electrical conductivities $\sigma_{lead}$ and $\sigma_{2DEG}$ of the leads and the 2DEGs, respectively \cite{Syme1989.jop.condensedmatter}. However, with $\kappa_{WF,lead} + \kappa_{WF,2DEG} < 0,01$ Wcm$^{-1}$K$^{-1}$ in the temperature range $T < 75$ K the thermal conductivities of the leads and the 2DEG are negligible compared to the measured thermal conductivities.\\

\section{Thermopower}
\begin{figure}[tb]
\centering
\includegraphics[width=0.5 \textwidth]{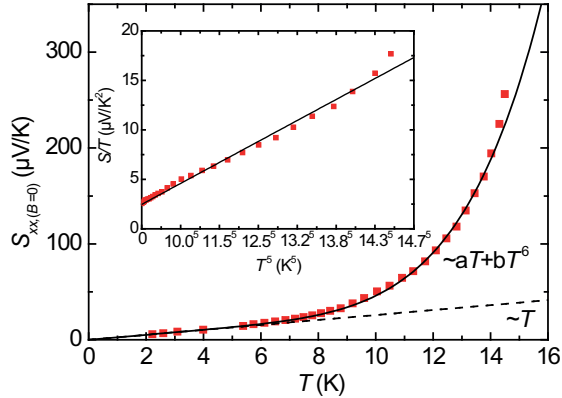}
\caption{(color online) Temperature dependence of the zero-field thermopower. The dashed line illustrates the diffusion-part of the thermopower $S_d$ calculated via the Mott formula with the experimental electron density and a power factor $p=-0.5$. The solid line represents the total thermopower calculated as $S=S_d+S_{ph}$. The inset shows a plot of $S/T$ vs. $T^5$ together with a fit $\propto T^5$. The intercept of the fit at $T=0$ is at 2.2 $\mu$V/K$^2$.}
\label{fig:Thermopower}
\end{figure}

Utilizing the low thermal conductivity of the membrane compared to bulk material, the thermopower $S$ was determined by measuring the thermovoltage between contacts that were only 34 $\mu$m apart. Here we report on measurements performed at temperatures between 2.7 K and 14 K without an external magnetic field. 
The thermopower $S=S_d+S_{ph}$ arises from two contributions, the charge carrier diffusion $S_d$ and the so-called phonon drag $S_{ph}$.\cite{Sankeshwar.pss,Ruf1998.prb} The latter is caused by the phonon wind, from which momentum is transferred to the charge carriers. To calculate the diffusion thermopower $S_d$, we assume the Mott formula\cite{Mott1969PhysRev,Nicholas.1985.JPhysC., Sankeshwar.pss} to be a good approximation for temperatures below 20 K: 
\begin{align}
	S_{d}=- \frac{\pi^2k^2_B}{3e}\frac{T}{E_F}(p-1) \label{zero-thermopower}.
\end{align}
The diffusion thermopower thus is proportional to the ratio of the temperature to the Fermi temperature of the electronic system. Furthermore, it depends on the parameter $p$ that takes account for the energy-dependence of the carrier scattering time. As in previous publications \cite{Sankeshwar.pss, basic-prop-of-semiconductors}, we assume that the electronic conductivity can be expressed in terms of an energy dependent relaxation time $\tau=\tau_0E^p$, where $E$ is the electron energy, $\tau_0$ a function of temperature (but independent of $E$) and the power factor $p$ is the parameter occurring in eq. (\ref{zero-thermopower}). The power factor in turn depends on the scattering mechanisms at work. \\
The dashed line in Fig. \ref{fig:Thermopower} is calculated with the experimental carrier density of $1.3 \cdot 10^{11}$ cm$^{-2}$ yielding a Fermi energy of 4.7 meV, which is related to a Fermi temperature of 55 K. 
The slope corresponds to a power factor of $p=-0.5$. In previous studies on AlGaAs-GaAs HEMT structures the factor $p$ was mostly found to be on the order of unity.\cite{Fletcher1994.prb, Cyca1992.JPCondMatter} It was associated with remote and background impurity scattering. However, the carrier densities were much larger in those studies compared to the carrier density in our sample. Different scattering mechanisms are dominant at different charge carrier densities $n$ so that $p$ will also depend on $n$. Karavolas and Butcher \cite{Karavolas1991.JPhysCondMatter} have calculated the parameter $p$ in the carrier density range from $5$ to $12\cdot 10^{11}$ cm$^{-2}$ for HEMT structures on bulk substrates. They considered interface roughness, remote and background impurities and found that in this carrier range $p$ can take values between -1.5 and 1.5. In our sample, scattering due to strain fields and even dislocations may occur after the suspension of the membrane so that additional scattering mechanisms influence the parameter. \\ 
In Hall bars on bulk GaAs substrates, the phonon drag dominates the thermopower already in the sub-Kelvin range.\cite{basic-prop-of-semiconductors, Fletcher1994.prb} In contrast, in our membrane the diffusion thermopower dominates up to much higher temperature. The temperature dependence of the phonon-drag thermopower $S_{ph}$ is generally taken to be of the form $S_{ph}\propto\Lambda T^n$, with $\Lambda$ being the phonon mean free path and an exponent $n$ that is 4 or 6 for electron-phonon coupling by piezoelectric or by deformation potential interaction, respectively.\cite{Kubakaddi2004.prb, Fletcher1997.prb, Sankeshwar.pss}\\ 
The solid line in Fig. \ref{fig:Thermopower} represents a fit to the total thermopower calculated as $S=S_d+S_{ph}=aT+bT^6$ with the parameter $a=2.47$~$\mu$V/K$^2$ obtained from eq. (\ref{zero-thermopower}). The best fit is obtained when we take a constant phonon mean free path, which yields the fitting parameter $b=2.12\cdot 10^{-5}$~$\mu$V/K$^7$. Interestingly, the calculation agrees even for temperatures higher than 10~K where the mean free path $\Lambda$ determined from the run of the thermal conductivity starts to become strongly $T$-dependent. We note that the introduction of an additional $T^4$ term to the fit results in a less accurate matching with the measured data. \\
In our suspended structure, the phonon-drag contribution starts to dominate the total thermopower beyond 7~K where the thermopower deviates from the linear run as obvious in Fig. \ref{fig:Thermopower}. The inset of Fig. \ref{fig:Thermopower} shows a plot of $S/T$ vs. $T^5$ together with a calculated fit proportional to $T^5$. The intercept at $T=0$ is at 2.2~$\mu$V/K$^2$ in good correspondence with the value calculated by the Mott formula.\\ 
In previous publications on thermopower studies with HEMTs on bulk GaAs crystals the phonon-drag sets in at much lower temperature. Furthermore, the temperature dependence is described by an exponent $3\leq n \leq 4$ instead of $n=6$.\cite{basic-prop-of-semiconductors, Tieke1995.prb} From this it was concluded that phonon drag is dominated by piezoelectric phonon-electron coupling in HEMTs on bulk GaAs. The $n=6$ power law observed in our HEMTs on thin membranes thus might indicate a different coupling mechanism. We note, however, that for a quantitative analysis it must be considered that the contact separation is on the scale of the phonon mfp. In all previous works where lattice heating was applied to generate a thermal gradient \cite{basic-prop-of-semiconductors,Fletcher1994.prb, Ruf1998.prb, Ying1994.prl, Cyca1992.JPCondMatter, Chickering2009.PRL} the devices were macro scaled with lengths that were several times the phonon mean free path. This indicates that the small contact separation is responsible for the fact that the phonon-drag signal in our device is much smaller than in macroscopic samples of previous studies.\\

\section{Conclusion}
In conclusion, we have used a suspended GaAs membrane to establish strong temperature gradients along a micro-scaled Hall bar containing a 2DEG. From the temperature gradient the thermal conductivity of the membrane was determined. It is found to be up to two orders of magnitude smaller than GaAs bulk values. Furthermore, the run is similar to data for micro-scaled GaAs structures previously reported.\cite{Fon2002.prb} The zero-field thermopower of the low-dimensional electron gas is found to be dominated by thermo-diffusion in the low temperature range up to 7 K. This temperature is significantly larger than previously reported values for non-suspended HEMT structures. The phonon-drag thermopower is found to be strongly suppressed with respect to HEMTs on bulk GaAs substrates, and an $n=6$ power  law is observed for the temperature dependence. We attribute the observations to the small dimension of the membrane in combination with the small distance between the contacts of the device. The $n=6$ power law might indicate a different electron-phonon coupling in the suspended membrane as compared to bulk samples. The small contact separation, which is of the order of the phonon mfp, leads to a suppression of the phonon-drag signal. A theoretical model is needed to reveal how the small contact separation influences the temperature dependence of the phonon-drag contribution to the thermopower.

\section*{Acknowledgments}
The authors thank the Deutsche Forschungsgemeinschaft for financial support via SPP 1386 "Nanostructured Thermoelectrics".

\end{document}